\documentclass[aps,prl,twocolumn,superscriptaddress,showpacs]{revtex4}

\usepackage{amsfonts}
\usepackage{graphicx}
\begin{document}
\title{Spin-$\frac{1}{2}$ fermions on spin-dependent optical lattices}
\author{M. ~A. Cazalilla}
\affiliation{Donostia Int'l Physics Center (DIPC), Manuel de Lardizabal,
4. 20018-Donostia, Spain.}
\author{A.~F. Ho}
\affiliation{School of Physics and Astronomy,  The University of Birmingham,
Edgbaston, Birmingham B15 2TT, UK.}
\author{T. Giamarchi}
\affiliation{University of Geneva, 24 Quai Enerst-Ansermet,
CH-1211 Geneva 4, Switzerland.}
\begin{abstract}
We study the phase diagram of one dimensional spin-$\frac{1}{2}$
fermionic cold atoms. The two ``spin'' species can have  different
hopping or mass. The phase diagram at equal densities of the species
is found to be very rich, containing  Mott insulators and superfluids. 
We also briefly discuss coupling  1D systems
together, and  some experimental signatures of these phases. In
particular, we compute the spin structure factor for small momentum, which should allow  the spin gap to be detected.
\end{abstract}
\date{\today}
\maketitle

Quantum engineering~\cite{LWZ04,HCZ02} 
of strongly correlated many-body systems has recently become
possible thanks to the spectacular advances in trapping ultracold
atoms in optical lattices~\cite{G02,M03,S04,K04} or in
microchip traps~\cite{Reichel}.  This has led to the study of
models that would otherwise be hard to realize
in solids, which may shed light on basic issues in 
quantum many body physics,   including
the understanding of \emph{e.g.} the origin
of high-T$_c$ superconductivity in doped copper oxides.
In particular, correlated  boson~\cite{G02,HCG04,DG04,S04},
Bose-Fermi~\cite{CH03}, 
and Fermi~\cite{FS04,FRZ04,GRJ03} systems have
received much experimental and 
theoretical attention in recent times.

In sharp contrast to electrons in solids, in cold atomic  systems,
different types of atoms (different hyperfine states or different
atomic species) can be trapped and controlled \emph{independently},
such that the  hopping, strength and sign of interactions
(inter- or intra-species) and densities can be \emph{continuously}
tuned. For example, Mandel \emph{et al.}~\cite{M03} managed to control
independently the periodic potential  for each atom
type loaded in an optical lattice. A p-wave  
Feshbach resonance~\cite{Sa04} can create a tunable
asymmetry in the interactions in a multi-species
Fermi gas. All this, of course, leads to a much 
richer physics, which remains to be understood.

\begin{figure}[t]
\begin{center}
\includegraphics[width=\columnwidth]{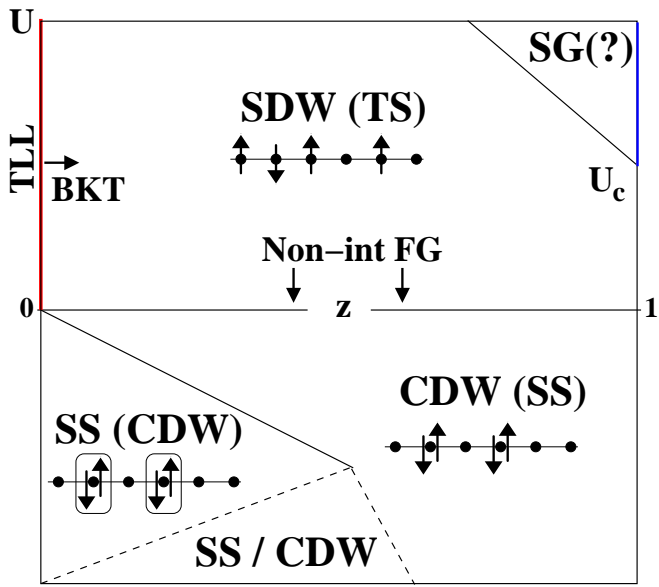}
\caption{Schematic phase diagram for the model in Eq. (\ref{Ham})
with equal number of spin up and down fermions
away from half-filling (\emph{i.e.} $N_{\uparrow 0} = N_{\downarrow 0} \neq M/2$) . 
The interaction strength is $U$ and  
$z = |t_{\uparrow} - t_{\downarrow}|/
(t_{\uparrow}+t_{\downarrow})$. All phases (SDW: spin density
wave, CDW: charge density wave, SS singlet superfluid, TS triplet
superfluid)   exhibit a spin gap $\Delta_s$ 
(however, $\Delta_s = 0$ for $U = 0$  and $z = 0$ with $U > 0$.
A cartoon of the type of order characterizing
each phase is also shown. In the area between dashed lines
the dominant order (either CDW or SS) depends on the lattice filling
(see text for more detailed explanations). In the SG phase, 
spin up and down fermions are segregated (demixed). 
\label{fig1}
}
\end{center}
\end{figure}

In this paper, motivated by these recent developments and the
availability (now or soon) of fermions in elongated
traps~\cite{K04,Reichel}, we study the interesting effects of having
different Fermi velocities  for two species of fermions in
one dimension (1D). With equal densities of the two species, this
system is  different from the case of a two-leg
spinless ladder~\cite{LH00,G04} or  the spin-$\frac{1}{2}$ 
electrons in a magnetic field~\cite{PS93,G04}. One main result 
of this paper is the phase diagram
as a function of velocity difference, for equal densities
(Fig.~\ref{fig1}). With repulsive interactions, a finite velocity
difference breaks the SU(2) spin symmetry 
and turns the gapless Tomonaga-Luttinger
liquid (TLL) into an Ising 
spin-density wave with a spin gap.  Demixing may occur if
one type of fermions has a very tiny velocity. With attractive
interactions, a singlet superfluid (SS) of bound pairs of fermions
of different types may give way to a charge density wave (CDW) of
pairs for sufficient velocity difference. We also briefly study the
effects of  a small tunneling term  coupling an array of 1D tubes
together. In particular, if there are different densities of
fermions in neighboring tubes, a triplet superfluid (TS) may become
stable for \emph{ repulsive} interactions. Finally, we describe how
to detect the spin gap in these phases by measuring the dynamical
spin structure factor, which we have computed. In contrast to
previous studies~\cite{FDL95,L04}, we have worked out 
the phase diagram  for equal number  of spin up and down 
fermions  as a function of the (Fermi) velocity difference
and considered coupling the 1D systems together.

We study the following  generalized Hubbard model:
\begin{equation}
H = -\sum_{\sigma, m}
t_{\sigma}   \left( c^{\dag}_{\sigma  m} c_{\sigma  m+1} + {\rm H.c.} \right)
+  U\sum_{m} n_{\uparrow  m}
n_{\downarrow  m}.
\label{Ham}
\end{equation}
This Hamiltonian describes a 1D Fermi gas with contact
interactions (related to $U$) \emph{prepared} with $N_{0\sigma}$ fermions
(i.e. we work in the \emph{canonical} emsemble) and loaded in a 1D
optical lattice with $m=1,\dots, M$ 
lattice sites; $n_{\sigma m} = c^{\dag}_{\sigma m} c_{\sigma m}$, and 
$\sigma = \uparrow, \downarrow$ is the spin index that may refer 
to two hyperfine states, or to {\it e.g.} 
$^6$Li and $^{40}$K. Even though there may be no \emph{true} spin
symmetry, we will continue to use the spin (and magnetic) language
to describe this binary mixture. 
We assume the number of fermions of
each spin species is \emph{separately} conserved, \emph{ i.e.} one
spin type cannot be converted to another. Motivated by the
experimental considerations above, we allow for different hopping
$t_{\sigma}$ for different spins. This Hamiltonian may be realized
in either a quasi-1D chip trap \cite{Reichel} or in a 2D optical
lattice, which is made up of an array  of 1D gas
tubes~\cite{S04,HCG04} weakly coupled by a hopping  $t_{\perp} \ll
\min\{t_{\uparrow}, t_{\downarrow} \}$. When the finite ``charging
energy'' (due to the finite length of each tube) exceeds the renormalized
hopping $E_{J} \propto t_{\perp}$~\cite{HCG04,unpub}, the tubes are 
decoupled from one another and a set of independent 1D tubes is 
recovered~\cite{HCG04}. Although we assume there is a (spin-dependent) 
periodic potential parallel to the tubes such that (\ref{Ham}) applies, 
much of what is discussed below also applies 
in the absence of this potential when the two species have
different masses.  More discussion on engineering  Hamiltonians like
(\ref{Ham}) can be found in~\cite{LWZ04}.  

We first study  the homogeneous 1D system 
in the thermodynamic limit. Finite-size and trap effects 
will be discussed below in  connection with possible experiments.

 The weak coupling limit $|U| \ll
\min\{t_{\uparrow},t_{\downarrow}\}$ can be solved by taking the
continuum limit of (\ref{Ham}) in the standard way \cite{G04} and
linearize the dispersion around the Fermi points $\pm k^{\sigma}_F =
\pi N_{0\sigma}/M a$, ($M$  is the number of  lattice sites in the
tube). This leads to the so-called ``g-ology'' representation~\cite{G04} with a
finite number of coupling constants  representing low energy
scattering processes. The coupling  $g^{\sigma}_{2||}$
($g_{2\perp}$) is the scattering amplitude for processes where a
small momentum $q$ is exchanged between fermions of equal (opposite)
spin at opposite Fermi points, for arbitrary  values of $k^{\sigma}_{F}$.
On the other hand,  $g_{1\perp}$  is the back-scattering
amplitude  where two fermions of opposite spin exchange  a momentum
$q \approx 2 k_F = 2 k^{\uparrow}_F = 2 k^{\downarrow}_{F}$, and is
relevant only when $N_{0\uparrow}=N_{0\downarrow}$; $g_{3\perp}$ is
the amplitude for \emph{umklapp} scattering ($q \approx
2k^{\uparrow}_F + 2 k^{\downarrow}_F  =\pi/a$) and is important only
at \emph{half-filling}: $N_{0\uparrow} + N_{0\downarrow} = M$. Thus
for generic fillings, $g_{1\perp}$ and $g_{3\perp}$ are irrelevant,
and the system is a TLL~\cite{G04}), which has a
completely gapless spectrum of two distinct branches of phonons.

We focus here on the case $N_{0\uparrow} = N_{0 \downarrow} \neq M/2$. 
The case of a half filled lattice $N_{0\uparrow} =  N_{0\downarrow} = M/2$ is more
involved and will be reported elsewhere \cite{unpub}

 The physical properties can be
established by analyzing the renormalization group (RG) flow of the
various scattering amplitudes upon the varying of a cutoff such as
the temperature $T$. To second order in the interaction
parameters, the RG flow is \cite{unpub}:
\begin{eqnarray}
\dot{y}^{\sigma}_{2||} &=& r_{-\sigma} y^{2}_{1\perp}, \quad \quad
\dot{y}_{2\perp} = -y^2_{1\perp},\label{RG1}\\
\dot{y}_{1\perp} &=& \left(r_{\uparrow} y^{\uparrow}_{2||} +
r_{\downarrow} y^{\downarrow}_{2||} - 2 y_{2\perp} \right) y_{1\perp},\label{RG2}
\end{eqnarray}
where  $y_{\alpha} = g_{\alpha}/\pi \hbar v$ are dimensionless
couplings,  $v = (v_{\uparrow} + v_{\downarrow})/2$ the mean
velocity and $r_{\sigma} = v/2v_{\sigma}$; $\dot{y}_{\alpha} =
dy_{\alpha}/d\ell$, with $\ell =  \ln (\Lambda/T)$. Eqs.
(\ref{RG1},\ref{RG2}) can be mapped to the RG equations of the
Berezinskii-Kosterlitz- Thouless (BKT) transition in terms of
$y_{2s} =  -\sum_{\sigma }r_{\sigma} y^{\sigma}_{2||}+ 2 y_{2\perp}$
and $y_{1\perp}$. The behavior of the BKT equations  is entirely
determined \cite{G04} by the constant of motion ${\cal
C}=y^{2}_{1\perp}  - y^2_{2s}/2(r_{\uparrow} r_{\downarrow} + 1) =
(Ua/\hbar v)^2 z^2/(2-z^2)$, where $z = |t_{\uparrow} - t_{\downarrow}|/(
t_{\uparrow} + t_{\downarrow})$ is the key velocity difference
parameter. For $z = 0$ we recover the well-known results for the
spin-symmetric Hubbard model~\cite{G04}. However, for $z\neq 0$
and $U\neq 0$, ${\cal C} > 0$, the scattering amplitude
$y_{1\perp}(\ell)$ diverges as the system is cooled down to its
ground state.  This signals the formation of bound states, and the
opening of a gap in the spin sector (the charge excitations remain
gapless).  For  $z \ll 1$, the gap has thus the characteristic BKT
form $\Delta_s \sim \Lambda\: e^{-A/\sqrt{\cal C}} \simeq \Lambda \:
e^{-A'/|t_{\uparrow} - t_{\downarrow}|}$, where $\Lambda$ is of the
order of $t_{\uparrow} \approx t_{\downarrow}$ and $A, A'$ are
constants. Note that this gap is non-perturbative in $|t_{\uparrow}
- t_{\downarrow}|$.

The  properties of the  spin-gapped phase depend on the sign of $U$.
Ground states of 1D systems are characterized by the dominant form
of order that they exhibit, which is typically quasi-long range in
character, true long-range order  being only possible in 1D when a
\emph{discrete} symmetry is broken. For $U > 0$ and $z\neq 0$,
then $y_{1\perp}(\ell) \to \infty$, and a bosonization study \cite{unpub}
shows that  the dominant order is a spin-density wave (SDW) and  the
subdominant order is triplet superfluidity  (TS). In the
attractive case ($U < 0$), as  $z$ is increased, the dominant
order  changes from a singlet superfluid phase (SS)  to  a
charge density wave (CDW), with CDW and SS being the
subdominant order in the former and latter case, respectively. We
wish to point out that our analysis takes fully into account the
marginal (in the RG sense) coupling between the gapless charge and
the gapped spin modes arising at $z\neq 0$, which leads to an often
substantial decrease in  the value of the Luttinger-liquid parameter
$K_c$ (proportional to the charge compressibility). In particular,
for $U< 0$ we find that $K_c$ goes from $K_c>1$ to $K_c<1$ as $z$
is increased, which changes the character of the dominant
correlations from SS to CDW, as described above. A summary of the
phase diagram is shown in Fig.~\ref{fig1}.

The weak-coupling regime smoothly crosses over to the strong
coupling regime $|U| \gg \max\{t_{\uparrow}, t_{\downarrow} \}$, as
confirmed by a strong coupling expansion of (\ref{Ham}). We only
give here the main steps, technical details can be found in
\cite{unpub}. It is simplest to first consider a half-filled lattice
with $N_{\uparrow 0} = N_{\downarrow 0} = M/2$. For the strongly
repulsive case $U \gg \max\{t_{\uparrow}, t_{\downarrow} \}$
fermions cannot hop around and there is a gap of order $U$ to charge
excitations. Degenerate perturbation theory~\cite{T77} then shows that in this limit
the Hamiltonian in (\ref{Ham}) maps to the Heisenberg-Ising (XXZ)
spin chain $H_{\rm XXZ} = J \sum_{m} \left[{\bf  S}_{m}
\cdot {\bf S}_{m+1} + \gamma \, S^{z}_{m}
S^{z}_{m+1} \right]$, where the ${\bf S}_{m}$ denotes the spin operator
at site $m$, $J = 4 t_{\uparrow} t_{\downarrow}/U$, and the anisotropy $\gamma =
(t_{\uparrow} - t_{\downarrow})^2/2t_{\uparrow}t_{\downarrow}
\propto z^2$. Thus, for unequal hopping ($z > 0$), the
chain is in the Neel phase (SDW with true long-range order),
and has a spin gap which for small $z$ is
 $\Delta_s \sim J\:  e^{-A''/\sqrt{\gamma}} = J\:
e^{-A'''/|t_{\uparrow}-t_{\downarrow}|}$~\cite{FDL95}. Note the same
non-perturbative dependence on $t_{\uparrow}-t_{\downarrow}$ as for
the weak coupling regime. Away from half-filling, the system is
described by a $t$-$J$-like model with anisotropic spin interactions.
The charge gap is destroyed ($K_c  = \frac{1}{2}$ close to
half-filling~\cite{G04}), but the spin gap remains and the dominant
order is still SDW. Physically, the finite velocity difference
breaks the SU$(2)$ spin symmetry to the lower Z$_2 \times$ U$(1)$.
Thus, the TLL becomes an Ising anti-ferromagnet in the spin sector. 

 For $U\ll 0$, 
degenerate perturbation theory shows that (\ref{Ham}) is   equivalent
to a model of tightly bound fermion pairs (hard-core bosons)
annihilated by $b_{m} = c_{\uparrow m} c_{\downarrow m}$. 
Their hopping amplitude  is $J = 4t_{\uparrow}
t_{\downarrow}/|U|$, and they interact with strength $V =  2
(t^{2}_{\uparrow} + t^2_{\downarrow})/|U|$ when sitting at
nearest-neighbor sites. This model can be mapped to the above XXZ chain 
via $b_{m} \to S^{-}_{m}$ and $\left(b^{\dag}_{m} 
b_{m} - \frac{1}{2} \right) \to S^{z}_{m}$.
At half-filling, charge excitations are gapless for equal hopping
and SS is the dominant order~\cite{G04}. However, with unequal
hopping the spectrum of the tube is fully gapped, becoming a CDW
with true long-range order, a spin gap of order $|U|$ (energy
to break a pair), and a charge gap $\Delta_c \sim J \:
e^{-A/|t_{\uparrow} - t_{\downarrow}|}$. Away from half-filling, the
spin gap remains $\sim |U|$ but the bosons are able to hop
(\emph{i.e.} the charge gap disappears). It is worth pointing out
that very close to half-filling for $z\neq 0$, the dominant order is
CDW since $K_c \to \frac{1}{2}$, as can be inferred from the exact
solution of the XXZ chain~\cite{H80,G04}. However, as the filling
deviates more and more from half-filling, $K_c$ rises above one and
the system becomes a 1D superfluid (SS). This change in the
character of the dominant order also takes place at constant
filling, provided the system is sufficiently far from half-filling:
a SS  ($K_c > 1$) can turn into a CDW ($K_c < 1$) as $|z|$ is varied
at strong coupling. This agrees with the  above 
weak coupling analysis. Note that at very low density 
($N_{0\sigma}/M\rightarrow 0$), and at least for not too different velocities, 
only a SS phase is possible: in this limit, (\ref{Ham}) maps to a
continuum (Gaudin-Yang like) model of interacting fermions with
spin-dependent mass. For $U\to -\infty$, the fermions  pair up to become a
1D superfluid (SS) of tightly bound pairs with irrelevant residual
interactions between the pairs.

Finally both the weak and strong coupling analysis described above
break down for sufficiently large $|t_{\uparrow} - t_{\downarrow}|$.
For weak coupling, linearization of the free fermion dispersion is
no longer justified  when $t_{\uparrow} \gg t_{\downarrow}$ (or
viceversa), that is, for $z \to 1$. In the large $|U|$ limit
degenerate perturbation theory becomes quite subtle. Unfortunately,
rigorous results are available only for $t_{\uparrow} = 0$ or
$t_{\downarrow} = 0$ ($z = 1)$, which is the limit of the
Falicov-Kimball model. In 1D, Lemberger~\cite{L92} (see also
\cite{FLX}) has proved that spin up fermions segregates from spin
down ones  for $U > U_c > 0$ at equal densities. There is no
segregation for $U<0$ at equal densities. As argued in
\cite{FLX,unpub}, it is quite likely that this segregated phase will
survive also when $|z|$ is not one but close to one.

The predicted phase diagram of Fig.~\ref{fig1} for a single 1D tube 
can be directly tested experimentally in cold atoms. However, it is 
also interesting to analyze the case when
the tubes are weakly coupled by  tunneling between the tubes.
We thus briefly describe the phase diagram for an array of
coupled 1D tubes in a 2D optical lattice geometry~\cite{S04,HCG04}. The
Hamiltonian for each tube at site ${\bf R}$ of the 2D lattice is
as in  Eq.~(\ref{Ham}), with all fermion operators now carrying the $\bf R$
label. The hopping between the \emph{nearest}  neighbor tubes 
at ${\bf R}$ and ${\bf R}'$ is described 
by $H_{\perp} = - t_{\perp} \sum_{\langle{\bf R}, {\bf R}'
\rangle} \sum_{m, \sigma} c^{\dag}_{\sigma {\bf R} m} c_{\sigma {\bf
R'} m}$, where $t_{\perp} \ll \min\{t_{\uparrow}, t_{\downarrow}
\}$, but such that fermions can now overcome the `charging energy' of the
finite-size tubes.  In general, when  the isolated tube has a gap 
 $\Delta_s \ll t_{\perp}$, $H_{\perp}$ is a relevant
perturbation (in the RG sense) that leads to coherent hopping of
fermions from tube to tube. Thus the ground state will  most likely be a
very anisotropic 3D Fermi liquid, which in turn may become unstable
to 3D CDW/SDW  formation or 3D BCS superfluidity under appropriate
conditions. This limit has been much studied in the past \emph{e.g.} in 
connection with organic superconductors (see~\cite{chem_rev} for a review). 
We shall not consider it here,
and instead we study
$t_{\uparrow} \neq t_{\downarrow}$  so that the gap $\Delta_s \gg t_{\perp}$.

Since the tubes (or at least a large number of them near the center
of the trap due to inhomogeneity effects) can develop a sufficiently large
spin gap as described above, coherent hopping between tubes is now
suppressed. However,  the term $H_{\perp}$ can generate,
through virtual transitions which are second order in $t_{\perp}$,
intertube interactions of two kinds~\cite{G04}: i)  particle-hole
pair hopping generates spin-spin and density-density
interactions: $H_1= \sum_{m,\langle {\bf R}, {\bf R}'\rangle} \left[
J_{\perp} \: {\bf S}_{{\bf R} m} \cdot {\bf S}_{{\bf R}' m} +
V_{\perp} n_{m {\bf R}} n_{m {\bf R}'} \right]$, and ii) fermion pair hopping
yields $H_2 = J_{c \perp} \sum_{m, \langle {\bf R}, {\bf R}'\rangle}
b^{\dag}_{{\bf R} m} b_{{\bf R} m+j}$, where $V_{\perp},
J_{c\perp}\sim t^2_{\perp}/\Delta_s$, 
$b_{{\bf R}m} = c^{\dag}_{\uparrow {\bf R} m} c_{\downarrow {\bf R} m}$ 
and $ja  \lesssim  \xi_s$ a distance smaller 
than the spin correlation length,  $\xi_s \propto \Delta^{-1}_s$.
The dominant term then drives a phase transition to a 3D ordered
phase~\cite{unpub}:  for the $U < 0$ case, if the
tubes are in the SS phase, then the dominant process is  
fermion pair tunneling, and the tubes  develop 3D
long-range superfluid order. The low-temperature properties of this
system become identical to the superfluid of bosons studied in
\cite{HCG04}. However, if the tubes are in the CDW ($U < 0$ and
sufficiently large $z$) or in the SDW ($U>0$) phases, the dominant
interactions arise from hopping of particle-hole pairs and lead to
insulating phases that are either 3D  CDW or SDW.  The ordering
temperatures in all cases (at small $t_{\perp}$) are power-laws:
$T_c \propto \Delta_s (t_{\perp}/\Delta_s)^{\alpha}$, with $\alpha^{-1} =
2(2-d)$ and $d$ the scaling dimension of the  dominant
inter-tube interaction. Interestingly, the SDW or CDW ordering is
anisotropic: incommensurate (relative to the optical lattice) along
the tube, but commensurate perpendicular to the tube direction.

Particle-hole hopping may drive a  transition to a 3D insulating
state with density wave order \emph{only if} the density in
neighboring tubes are equal or very similar: for a particle and a
hole to hop coherently at low-temperatures, they must be extracted
from opposite Fermi points of one tube and must match the momenta in
the neighboring tube by momentum conservation. If  the mismatch in
the density between tubes is sufficiently large, particle-hole hopping is
suppressed and only the hopping of fermion pairs (which carry zero net 
momentum) is possible. The system will then order as a
superfluid. Interestingly, for $U > 0$, TS is the subdominant
order in the spin gapped phase of the tubes. Suppression of
particle-hole pair hopping may then lead to a 3D triplet superfluid.
We note that  $T_c$ for these cases is also a power law of
$t_{\perp}/\Delta_s$.

The phase diagram shown in Fig.~\ref{fig1} holds, strictly speaking,
in the thermodynamic limit. In real 2D optical lattices only a  finite
number of fermions ($\sim 10^5$~\cite{K04})  can be loaded, but we expect all
predicted phases to appear. Due to the finite size, the phase
boundaries will not correspond to true phase transitions, but rather
to sharp crossovers. The trap can lead to phase coexistence and even to
suppression of quantum criticality~\cite{W04}, but we are concerned
here with the phases themselves and not with the quantum critical
points between them.

The most important signature of the single tube phases that we
predict is the existence of a spin gap, $\Delta_s$.    By measuring the absorption  of a
laser  that causes Raman transitions between the two hyperfine states
$\sigma =\uparrow, \downarrow$, it should be possible to measure~\cite{BZZ04}  the small momentum limit of the dynamic structure factor $S_s({\bf q}, \omega)$, which is the  Fourier transform of $S_s({\bf r},{\bf r}', t) = 
\langle S^{+}({\bf r}, t) S^{-}({\bf r}', 0) \rangle$.  Using the
so-called form factor approach~\cite{LZ01} ,
 we find~\cite{unpub} that at $T \ll \Delta_s$, the structure factor rises from zero as
$\sqrt{(\hbar \omega)^2 - (2\Delta_s)^2}$ for $\hbar\omega \geq 2\Delta_s$.

Concerning the coupled tubes, the most exotic phase is of course
the triplet superfluid (TS). To `engineer' it, we need to suppress
particle-hole hopping by making the number of fermions in neighboring
tubes sufficiently different. This could be achieved by imposing 
a rapid spatial variation of the
trap potential, or better, by means of a biperiodic optical potential in the
direction perpendicular to the tubes. The coherence  properties of
the 3D superfluid phases could be probed by exciting low frequency
collective modes in the transverse direction to  the tubes. Coherent
oscillations should exist only in the superfluid phases and not in
the  insulators.

We acknowledge useful conversations with T. Esslinger and his group, M. Fabrizio,
J. Fjaerestad, A. Georges, M. Gunn, D. Haldane, B.
Marston, M. Long, A. Nersesyan, B. Normand and A. Tsvelik. M.A.C.
is supported by \emph{Gipuzkoako Foru Aldundia}, A.F.H. by EPSRC(UK)
and DIPC (Spain), and T.G. by the Swiss National Science Foundation
under MANEP and Division II.

\end{document}